# Mode Clustering Based Dynamic Equivalent Modeling of Wind Farm for Small-Signal Stability Analysis

Xiuqiang He, *Student Member, IEEE*, Hua Geng, *Senior Member, IEEE*, and Geng Yang, *Senior Member, IEEE*

***Abstract*—Dynamic equivalent models (DEMs) are necessities for the small-signal stability (SSS) analysis of the power system with large-scale wind farms (WFs). This paper proposes a mode clustering based dynamic equivalent modeling method of WFs for the SSS analysis. It is deemed that a DEM can be used to represent the whole WF to evaluate its impact on the SSS of power systems, as long as the frequency response of the DEM adequately matches that of the detailed WF model around the frequency of oscillation modes of concern. Focusing on the concerned oscillation modes in the small-signal model of the whole WF, closely distributed modes of them on the complex plane are classified into a cluster and then represented by a single mode. By the linear superposition principle, the modal participation factor (MPF) regarding each of modal clusters are superimposed, generating a feature vector for each wind turbine (WT). Based on the feature vectors, the oscillation feature similarity of the WTs as well as their aggregation can be evaluated and performed easily. Taking the DC-link voltage control (DVC) mode of full-power WT converters as an example, the aggregated DEM is verified by frequency-domain analyses and time-domain simulations.**

## I. Introduction

WITH the ever-increasing penetration level of wind power generation into the power system, small-signal stability (SSS) issues of the power system induced by the integration of large-scale wind farms (WFs) have received much attention [1]. Unlike typical low-frequency oscillation phenomena in conventional power systems, wind turbines (WTs) often exhibit wide time-scale dynamics, and therefore emerging oscillations such as subsynchronous oscillation (SSO) introduced by WFs is with higher frequency than conventional low-frequency oscillations [2]. It is urgent to study new features of the emerging oscillation phenomena, especially interaction behaviors between WFs and the power system. In this regard, it is essential to firstly necessitate a simplified yet adequate WF model, which can exhibit the dynamics around the frequency of oscillation modes of concern [3]. It is acknowledged that detailed WF models are not applicable since their high complexity makes the stability evaluation and time-domain simulations difficult and impractical. Simplified dynamic equivalent models (DEMs) are often employed in previous studies [4]–[23].

One of the most typical methods to derive a DEM is based on WT clustering and aggregation, which actually draws lessons from the concept of the coherence-based equivalent of synchronous generators [24]. Taking into account the clustering index indicating dynamic behaviors of individual WTs, multiple WTs having similar dynamic behaviors (i.e., similar clustering indexes) are bracketed and then aggregated into a single one. The modeling process is concluded as follows.

1) STEP 1: WF configuration information and parameter collection; 2) STEP 2: Single WT modeling and simplification; 3) STEP 3: Detailed WF model construction; 4) STEP 4: WT clustering and aggregation; 5) STEP 5: Collector network aggregation; 6) STEP 6: Equivalent model parameter identification based on measurements at the point of interconnection (POI); 7) STEP 7: Model validation. Performing either STEP 1 or 6 for parameter determination corresponds to the certain parameter-based DEMs [4]–[20] or the identified parameter-based DEMs [21]–[23], respectively. The latter one is usually used to address the situations where WF configuration information or parameters are unknown or uncertain, and the relevant studies [21]–[23] mainly focus on the parameter identification techniques in STEP 6. Actually, maintaining accurate parameters is the responsibility of independent system operators (ISOs) [22]. From the perspective of transmission system operators (TSOs), the above certain parameter-based DEMs are much popular as long as accurate parameters can be obtained in advance. Prior studies [4]–[20] regarding the certain parameter based DEMs primarily focus on the WT classification issues in STEP 4, where various clustering indexes are adopted, e.g., the single-machine representation [4]–[6] under the same wind speed condition, the multi-machine representation (including the semi-aggregation [7]–[9] and full-aggregation [10], [11]) with different wind speeds. Also, generator rotor speed [12], protection trigger action after grid faults [13], stator short-circuit current curve [14], post-fault active power response curve [15], [16] are applied as the clustering indexes. Moreover, References [17] and [18] utilize the feature vector composed of multiple time-domain state variables as the clustering index. In [19] and [20], several probabilistic clustering methods considering wind velocity uncertainty in long time scales are developed.

To verify and evaluate the adequacy and applicability of existing DEMs, simulations under large disturbances are performed in [25]–[27]. The results in [27] show that there are inevitable deviations in the DEMs due to the non-uniformity among individual WTs, which have large impacts on transient stability prediction of the power system. While using wind speed as the clustering index, Reference [28] focuses on the SSS and analyzes the effect of the equivalent error on the oscillation modes within WFs. The finding indicates that the existing practice of DEMs is inadequate and it is quite neces-

The authors are with the Department of Automation, Tsinghua University, Beijing 100084, China (e-mail: he-xq16@mails.tsinghua.edu.cn; genghua@tsinghua.edu.cn; yanggeng@tsinghua.edu.cn).

sary to develop improved aggregation modeling methods [28].

As reported in [25]–[28], although the modeling process of DEMs has gained recognition, unfortunately, there is not a recognized DEM that can be used as a generic model for research on a particular type of stability such as the SSS. The authors contend that the main reason lies in that there is a great divergence between most of the existing DEMs [4]–[23] and the specific model requirements in the research on a particular type of stability issue. In [4]–[23], it is usually considered that a perfect DEM is almighty as far as possible, i.e., it is able to provide high accuracy of output characteristics under: 1) steady-state conditions, 2) wind speed fluctuations in long time scales, and 3) small and large grid disturbances in short time scales. However, it is commonly acknowledged that the capability of a certain desired model is determined by its specific requirement, which varies from different types of stability issues. Therefore, the authors believe that it is unpractical to satisfy a wide range of model requirements with only a narrowly specific DEM. This standpoint has been supported by several recent published papers [29]–[32], where particular application scenarios of DEMs are taken into consideration and therefore well-directed DEMs are proposed for the specific requirements, such as power system dispatch [29], automatic generation control [30], and frequency regulation [31], [32].

For the SSS analysis, there are few studies concentrated on DEMs of WFs (small-signal DEMs for short). References [33]–[36] make pioneer work in building small-signal DEMs. Details on model reduction, aggregation, and parameter identification are discussed in [33], [34], but the rationality of the adopted clustering index is not explained. In [35], the concept of phase motion equation is employed to perform a one-by-one WT aggregation method with respect to the DC-link voltage control time scale. Limited to a synchronous power controlled photovoltaic power plant, the quantitative conditions for receiving a single-machine representation are derived in [36].

This paper is devoted to studying small-signal DEMs. Prior DEMs [4]–[20] are traditionally based on time-domain variables (e.g., wind speed) as the clustering index. While used in the SSS analysis, a fatal flaw of the DEMs is that the time scale of chosen time-domain variables may fail to cover the frequency range of oscillation modes of interest. For example, wind speeds merely indicate steady-state operating points rather than dynamic behaviors of WTs. Therefore, it can be inferred that the DEMs using wind speed as the clustering index cannot completely reflect the small-signal dynamic behaviors of WFs. Another shortcoming of existing DEMs is that the similarity of individual WTs is merely evaluated from the perspective of individual ones rather than the whole WF. It has been reported that inevitable interactions occur among multiple WTs [37] and output characteristics of the whole WF are also affected by the distributed network [27], [28]. In short, it is crucial to choose or design a reasonable clustering index to indicate the dynamic behaviors of concern. Furthermore, it is necessary to evaluate the similarity of the motion components from the perspective of the whole WF system.

Considering the fact that small-signal models can be considered as special linear time-invariant (LTI) systems, a generic equivalent modeling method based on mode clustering is proposed for a general form of distributed LTI system in Section II. In the light of the principle of linear superposition, closely distributed modes of the whole system are bracketed considering that they have similar motion features. Afterwards, a concrete dynamic equivalent modeling method of WFs for the SSS analysis is developed in Section III. Section IV verifies the effectiveness of the method. Without loss of generality, Sections III and IV take the DC-link voltage control (DVC) mode [38] of full-power WT converters as the concerned mode to demonstrate the method.

## II. A Generic DEM of Distributed LTI Systems

For an LTI system, an oscillation mode shows a type of oscillation characteristics, including the oscillation frequency and damping factor. For a distributed LTI system consisting of $N$ parallel subsystems (SSs), as shown in Fig. 1, a heap of adjacent modes shows similar oscillation frequencies and damping features. According to the superposition principle of linear systems, similar modes can be classified and superimposed to simplify the system. In other words, it is desired to substitute a cluster of similar modes with a single mode. As depicted in Fig. 1(b), if the effect of the collector network is ignorable, then the $N$ SSs having the same oscillation modes can be aggregated into a single one with rescaled capacity. A rigorous clustering, superposition, and aggregation demonstration are given as follows, where the modes are not always adjacently located.

### A. Clustering and Superposition of Similar Motion Modes

It can be considered that the individual SSs in Fig. 1(a) share the same system structures, similar system parameters, and thus close state trajectories. For a single SS, denote the oscillation mode of concern is $\lambda$ and the related state variable is $x$. There is a heap of modes, denoted as the modal set $\boldsymbol{\lambda} = \{\lambda_1, \lambda_2, \ldots, \lambda_i, \ldots, \lambda_N\}$, in the interconnected system consisting of $N$ SSs, and the related state vector can be denoted as $\boldsymbol{x} = \{x_1, x_2, \ldots, x_k, \ldots, x_N\}$. Let $\boldsymbol{A}$ be the state matrix of the whole distributed LTI system, with the input $\boldsymbol{u}$ and output $\boldsymbol{y}$. For any mode $\lambda_i$, let $\boldsymbol{U}_i$ (column vector) be the right eigenvector of $\boldsymbol{A}$, satisfying $\boldsymbol{A}\boldsymbol{U}_i = \lambda_i\boldsymbol{U}_i$; let $\boldsymbol{V}_i$ (row vector) be the left eigenvector of $\boldsymbol{A}$, satisfying $\boldsymbol{V}_i\boldsymbol{A} = \lambda_i\boldsymbol{V}_i$. By the theory of linear systems, the zero-input response $x_k(t)$ of the $k$th state variable in $\boldsymbol{x}$ is given by

$$x_k(t) = \sum_{i=1}^{N} C_{ki} e^{\lambda_i t} + \sum_{j} D_{kj} e^{\lambda_j t} \approx \sum_{i=1}^{N} C_{ki} e^{\lambda_i t} \quad (1)$$

with the given initial value $\boldsymbol{X}(0)$ ($\boldsymbol{X}$ denotes all the state variables, including but not limited to $\boldsymbol{x}$.) The first term in (1) is the time-domain response of the modal set $\boldsymbol{\lambda}$ where $C_{ki} = \boldsymbol{V}_i\boldsymbol{X}(0)U_{ki}$, and $U_{ki}$ is the $k$th element of $\boldsymbol{U}_i$. The second term in (1) is the time-domain response of the rest of the system modes. It can be considered that the concerned modal set $\boldsymbol{\lambda}$ is the dominant mode whereas the rest of the modes are non-dominant. Therefore, the modal components in the second term converge to zero quickly and then can be approximated by zero.

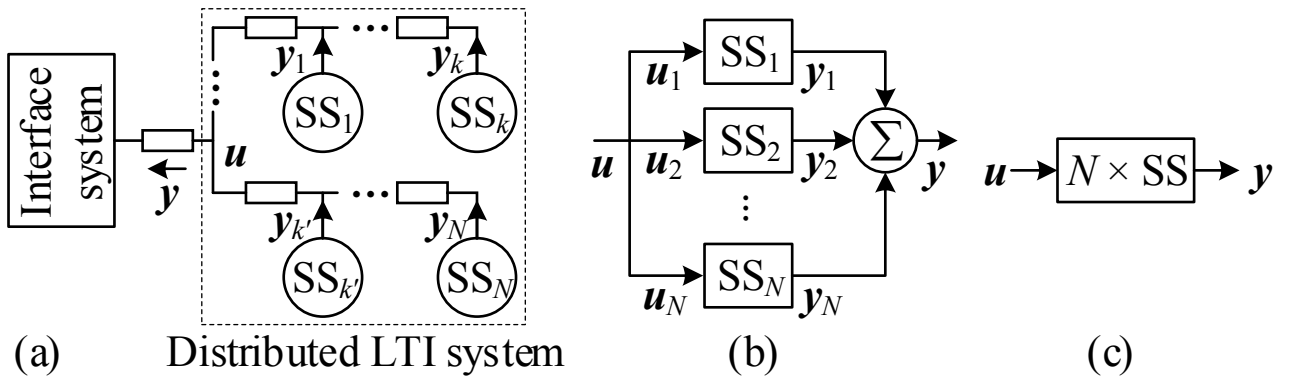

Fig. 1. Distributed LTI system consisting of $N$ parallel subsystems (SSs). (a) Complete system. (b) Simplified system after neglecting the effect of the collector network. (c) Aggregated system with $N$ identical modes.

TABLE I
MPFs OF THE DETAILED MODEL

| | $\lambda_1$ | $\lambda_2$ | … | $\lambda_i$ | … | $\lambda_N$ |
|---|---|---|---|---|---|---|
| $x_1$ | $f_{11}$ | $f_{12}$ | … | $f_{1i}$ | … | $f_{1N}$ |
| $x_2$ | $f_{21}$ | $f_{22}$ | … | $f_{2i}$ | … | $f_{2N}$ |
| … | … | … | … | … | … | … |
| $x_N$ | $f_{N1}$ | $f_{N2}$ | … | $f_{Ni}$ | … | $f_{NN}$ |

TABLE II
SUPERIMPOSED MPFs

| | $\boldsymbol{\lambda}^1$ | $\boldsymbol{\lambda}^2$ | … | $\boldsymbol{\lambda}^C$ |
|---|---|---|---|---|
| $x_1$ | $F_{11}$ | $F_{12}$ | … | $F_{1C}$ |
| $x_2$ | $F_{21}$ | $F_{22}$ | … | $F_{2C}$ |
| … | … | … | … | … |
| $x_N$ | $F_{N1}$ | $F_{N2}$ | … | $F_{NC}$ |

The modal participation factor (MPF) is used to evaluate the relevance between the state variable $x_k$ and the mode $\lambda_i$. Setting $x_k(0)$ as 1 whereas the other elements in the initial value $\boldsymbol{X}(0)$ as 0 gives rise to

$$x_k\left(t\right)=\sum_{i=1}^{N} f_{ki}e^{\lambda_i t}\text{, } f_{ki}=V_{ik}U_{ki} \tag{2}$$

where $V_{ik}$ is the $k$th element of $\boldsymbol{V}_i$, and $f_{ki}$ is exactly the MPF of $\lambda_i$ in $x_k$. Similarly, an MPF table is given in Table I.

In (1), if there are multiple modes located closely or identically on the complex plane, e.g., $\lambda_1$ overlaps with $\lambda_2$, then those modes can be merged, e.g., $e^{\lambda 1t} = e^{\lambda 2t}$. Classifying the elements in the modal set $\boldsymbol{\lambda}$ on the complex plane is able to act as an available solution to evaluate the similarity of the modes. Let the classification result be $\{\boldsymbol{\lambda}^1, \boldsymbol{\lambda}^2, \ldots, \boldsymbol{\lambda}^C\}$, which means that the $N$ modes are classified into $C$ clusters. Without loss of generality, it can be assumed that $\boldsymbol{\lambda}^1 = \{\lambda_1, \lambda_2, \ldots, \lambda_i\}$. Thus, it is expected to merge the modal components $\{e^{\lambda 1t}, e^{\lambda 2t}, \ldots, e^{\lambda it}\}$ in (1). In particular, it is also supposed to merge the modal components in (2), i.e., to superimpose each row of first $i$ columns in Table I, as given by

$$F_{k1}=\sum_{j=1}^{i} f_{kj}\text{, } k=1,\ldots,N. \tag{3}$$

Similarly, performing the superposition on the rest of $C-1$ modal clusters yields the superimposed MPFs, as given in Table II.

### B. Aggregation of Subsystems

From the global perspective of the whole distributed LTI system, the similarity of the motion components has been evaluated and similar motion modes have been merged and superimposed. Mapping the result of the modal superposition to the similarity of motion characteristics of the individual SSs (i.e., the individual elements of the state vector $\boldsymbol{x}$) is another crucial task to develop a DEM for the system. Fortunately, the superimposed MPFs (see Table 2) provide inherent mapping relations naturally as follows,

TABLE III
DYNAMIC EQUIVALENT MODELING PROCESS OF WFs

| |
|---|
| ✧ Single-WT modeling, simplification, and linearization |
| ✧ Linearized WF model construction |
| ✧ Modal clustering and superposition |
| ✧ WT and collector network aggregation |
| ✧ Model validation |

$$\text{SS}_k \Leftrightarrow x_k : \boldsymbol{F}_k=\begin{bmatrix}F_{k1} & F_{k2} & \cdots & F_{kC}\end{bmatrix}^T. \tag{4}$$

The vector $\boldsymbol{F}_k$ in (4) can be deemed as the feature vector of the subsystem $\text{SS}_k$. According to the concept of the MPF, the vector can not only indicate the dynamic characteristics of the subsystem participated by the modal set $\boldsymbol{\lambda}$, but also reflect the impact of the initial value of $\boldsymbol{x}$ on the degree of mode excitation. By observing and comparing the feature vectors of the subsystems, it is easy to identify the subsystems with similar oscillation characteristics, and then the subsystems can be aggregated to a single one with rescaled capacity.

## III. DEMs OF WFs

### A. Modeling Method and Process

Small-signal modeling of WFs can be conducted at some steady-state point: original nonlinear models are linearized at the operating point, resulting in small-signal models. Since small-signal models can be seen as LTI systems, the method presented in Section II can be specialized for WF modeling.

The modeling process is given in Table III. It is critical to firstly perform single-machine modeling for a particular application scenario, and also necessary model simplification to reduce the system order. After linearization of the single-machine model, the WF model can be assembled considering the collector network constraint. Based on the WF model, a specific DEM can be obtained following the method in Section II. Finally, the DEM is expected to be validated.

### B. Single-WT Modeling

For a clear and concrete demonstration, this study takes the DVC timescale dynamics (~10 Hz) of WT converters as the dynamics of concern, since it has been reported that the dynamics play an important role in multi-machine interactions and power oscillations occurring in actual WFs [37], [38]. Note that the modeling method is capable of any studied modes, not limited to the DVC modes discussed below. Moreover, it is known that most of the existing WTs still operate in the unity-power factor mode and the local reactive power compensation technique is commonly used, the modeled WTs in this study do not participate in the terminal voltage regulation.

Neglecting the current control dynamics (~100 Hz), the single-WT nonlinear and linearized models are obtained with the method in [37], [38], as shown in Fig. 2. The linearized model is expressed as follows,

$$\left(\Delta P_m-\Delta P_e\right)/\left(CU_{dc0}s\right)\left(k_{p2}+k_{i2}/s\right)=\Delta i_d \tag{5}$$

$$\Delta i_q=0 \tag{6}$$

$$\Delta u_q\left(k_{p2}+k_{i2}/s\right)1/s=\Delta\delta \tag{7}$$

where the power derivation $\Delta P_e$ is given by

$$\Delta P_e = \Delta u_d i_{d0} + u_{d0} \Delta i_d . \tag{8}$$

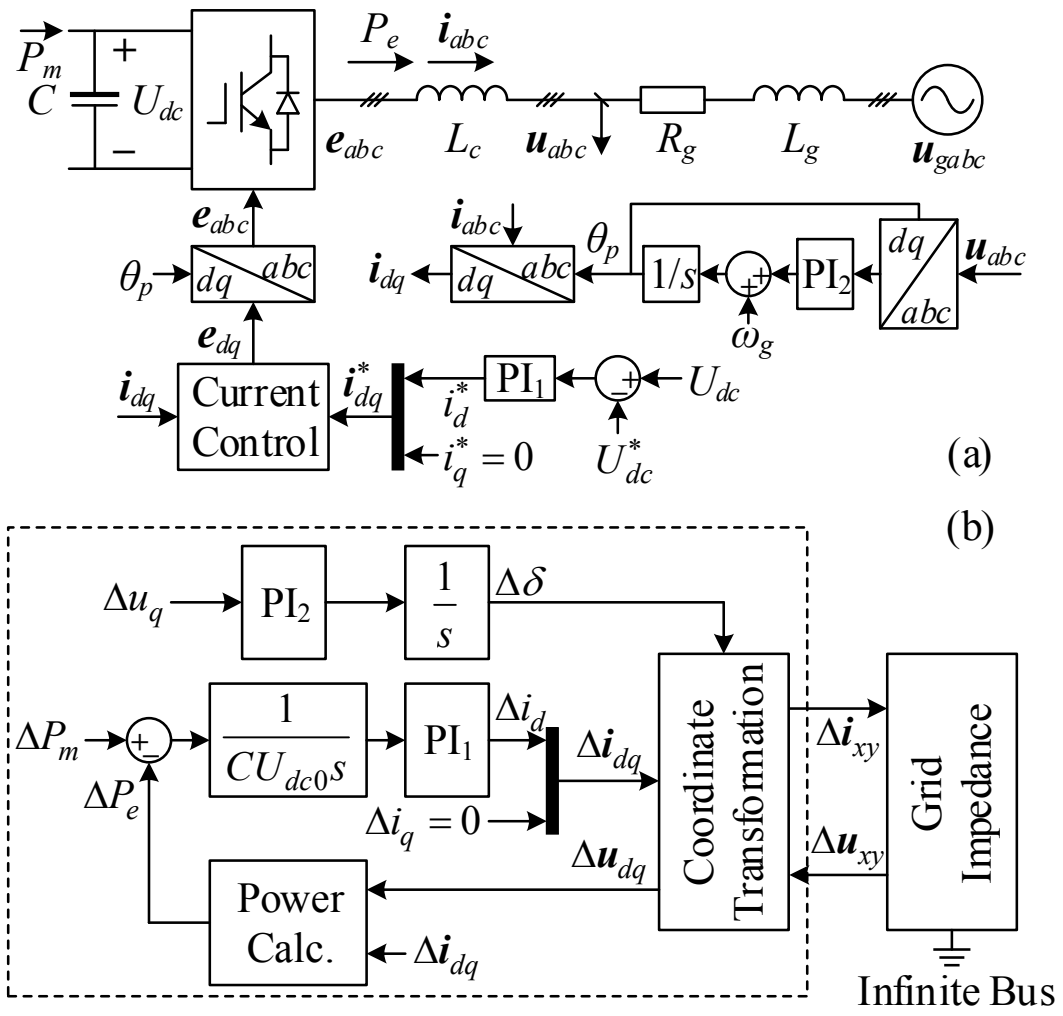

Fig. 2. Single-WT model. (a) Nonlinear model. (b) Linearized model.

The variables with subscript “0” represent the counterparts at the steady-state point (similarly hereinafter). It should be noted that (5)–(8) are expressed in the local phase-locked loop (PLL) reference frame. The coordinate transformation (let the included angle be $\delta$) between the PLL reference frame and infinite bus $XY$ reference with rotational frequency equaling the grid frequency $\omega_g$ can be expressed as

$$\begin{cases} \Delta \boldsymbol{u}_{dq} = \Delta \boldsymbol{T} \boldsymbol{u}_{xy0} + \boldsymbol{T}_0 \Delta \boldsymbol{u}_{xy} \\ \Delta \boldsymbol{i}_{xy} = \Delta \boldsymbol{T}^T \boldsymbol{i}_{dq0} + \boldsymbol{T}_0^T \Delta \boldsymbol{i}_{dq} \end{cases}, \boldsymbol{T} = \begin{bmatrix} \cos\delta & \sin\delta \\ -\sin\delta & \cos\delta \end{bmatrix} \tag{9}$$

where the superscript “$T$” denotes the transposition operator; $\boldsymbol{T}$ denotes the coordinate transformation matrix.

Thevenin equivalent grid model can be expressed as

$$\Delta \boldsymbol{u}_{xy} = \begin{bmatrix} R_g & -\omega_g L_g \\ \omega_g L_g & R_g \end{bmatrix} \Delta \boldsymbol{i}_{xy} . \tag{10}$$

### *C. WF Model Construction*

The WF model can be assembled with the individual WTs considering the collector network constraint. Prior to that, power flow calculation should be executed to determine the steady-state point of the entire system, where each WT is regarded as a $PQ$ node whereas the infinite bus is the sole $V\theta$ node.

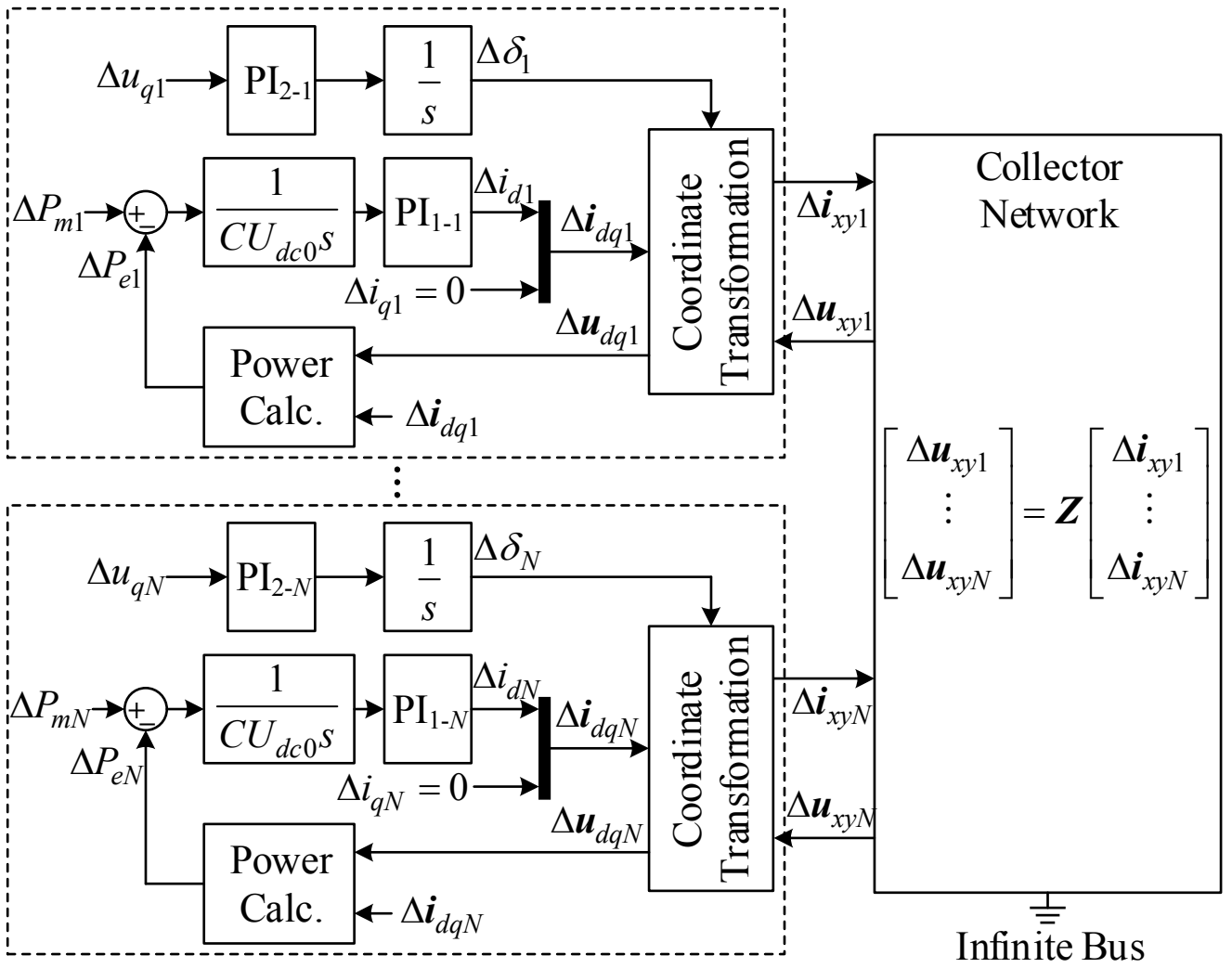

Fig. 3. Linearized WF model assembled with $N$ WTs.

Let the extended model with $N$ WT’s linearized models (shown in Fig. 3) be

$$\begin{cases} \Delta \dot{\boldsymbol{x}} = \boldsymbol{A} \Delta \boldsymbol{x} + \boldsymbol{B} \Delta \boldsymbol{u} \\ \Delta \boldsymbol{i} = \boldsymbol{C} \Delta \boldsymbol{x} + \boldsymbol{D} \Delta \boldsymbol{u} \end{cases} \tag{11}$$

where $\Delta \boldsymbol{u} = [\Delta \boldsymbol{u}_{xy1}, \ldots, \Delta \boldsymbol{u}_{xyN}]^T$ and $\Delta \boldsymbol{i} = [\Delta \boldsymbol{i}_{xy1}, \ldots, \Delta \boldsymbol{i}_{xyN}]^T$ denote the inputs and outputs of the $N$ WTs, respectively, and thus the other variables are self-explanatory.

The collector network equation is given by

$$\Delta \boldsymbol{u} = \boldsymbol{Z} \Delta \boldsymbol{i} \tag{12}$$

where $\boldsymbol{Z}$ is the node impedance matrix of the collector network, and (10) is a special case. From (11) and (12), it can be obtained

$$\Delta \dot{\boldsymbol{x}} = \left[ \boldsymbol{A} + \boldsymbol{B} \left( \boldsymbol{Y} - \boldsymbol{D} \right)^{-1} \boldsymbol{C} \right] \Delta \boldsymbol{x} = \boldsymbol{A}_s \Delta \boldsymbol{x} \tag{13}$$

where $\boldsymbol{Y} = \boldsymbol{Z}^{-1}$ is the node admittance matrix and $\boldsymbol{A}_s$ is the state matrix of the whole WF system.

### *D. Modal Clustering and Superposition*

After obtaining the state matrix $\boldsymbol{A}_s$, the DVC modes and the corresponding MPFs can be solved with the selective modal analysis [39], through which direct high-order calculation can be avoided. The modes can then be classified on the complex plane by a clustering algorithm such as the $k$-means algorithm. Thereafter, the feature vector for each WT can be formed, as illustrated in Section II.

### *E. WT Aggregation*

Based on the feature vectors, the WTs with similar motion features can be artificially identified easily since the differences among the feature vectors are quite significant (see Section IV). After that, the aggregation of the WTs is a straightforward process in the per-unit system [25], which is to rescale the capacity base to a certain multiple.

When it comes to the aggregation of the collector network, Reference [16] presents an analytical method to determine the equivalent parameters. It is reported that the method is analytical and has quite high precision, then is adopted in this study. So far, the small-signal DEM has been obtained.

### *F. Model Validation*

Since the DEM is based on the concept of using a single mode to represent a modal cluster, it is supposed for the DEM to provide sufficiently accurate modes to represent the original modes. In this regard, the modeling errors are defined by,

$$E = \max_i \left\{ \left| \lambda_i - \text{Centre}\left( \lambda_i \right) \right| / \left| \lambda_i \right| \right\} \tag{14}$$

$$E' = \max_i \left\{ \left| \lambda_i - \text{Nearest}\left( \lambda_i \right) \right| / \left| \lambda_i \right| \right\} . \tag{15}$$

In (14), Centre($\lambda_i$) denotes the centre of the cluster into which $\lambda_i$ is classified, and thus $E$ indicates the maximum relative error while using the cluster centre to represent individual modes. In (15), Nearest(Centre($\lambda_i$)) denotes the nearest mode in the DEM toward $\lambda_i$, and therefore $E'$ indicates the maximum relative error while using the nearest mode to represent individual modes.

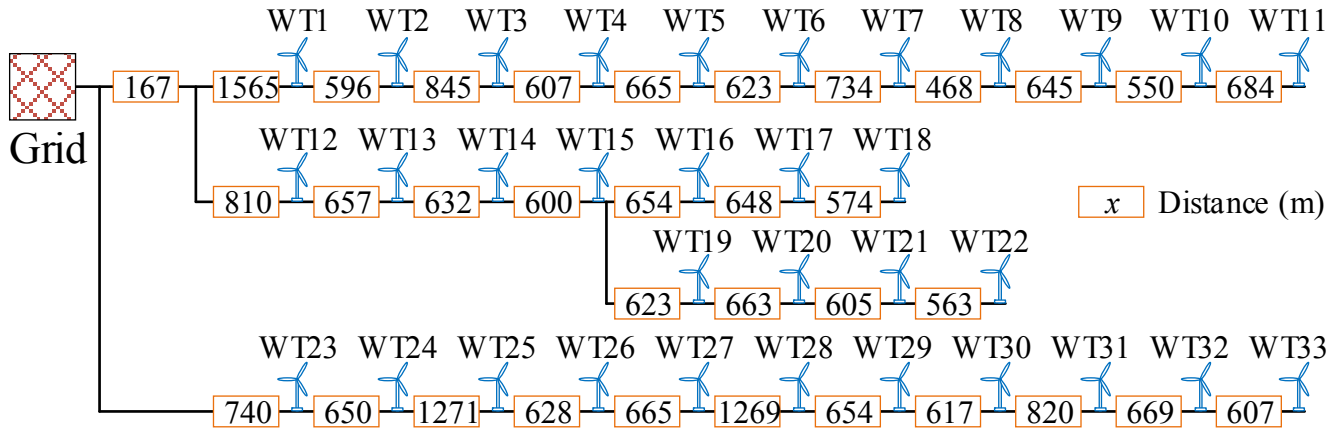

Fig. 4. Layout of an actual WF.

TABLE IV
WF PARAMETERS

| | $P_{m0}$ (p.u.) |
|---|---|
| WT01~WT11 | 1.00 1.00 0.95 0.95 0.95 0.90 0.90 0.85 0.85 0.80 0.80 |
| WT12~WT23 | 1.00 0.95 0.90 0.90 0.85 0.80 0.80 0.75 0.75 0.70 0.65 |
| WT24~WT33 | 1.00 0.90 0.85 0.85 0.80 0.70 0.65 0.65 0.60 0.50 0.45 |
| Line resistance | 0.1153 Ω/km |
| Line inductance | 1.05e–3 H/km |
| Capacity base | 1.5 MVA |
| Grid reactance $L_g$ | 0.01 p.u. |
| Grid resistance $R_g$ | 0.001 p.u. |
| DC-link capacitance $C$ | 90000 uF |

TABLE V
DVC PARAMETERS

| Case | $k_{p1}$ | $k_{i1}$ | WT Sequence Number |
|---|---|---|---|
| A | 1 | 300 | All |
| | 1 | 300 | 03, 08, 09, 12, 13, 15, 28, 29, 30, 31 |
| B | 2 | 300 | 05, 06, 07, 10, 14, 16, 17, 18, 19, 22, 23, 26, 27, 32 |
| | 3 | 300 | 01, 02, 04, 11, 20, 21, 24, 25, 33 |
| | 1 | 100 | 03, 08, 09, 12, 13, 15, 28, 29, 30, 31 |
| C | 1 | 300 | 05, 06, 07, 10, 14, 16, 17, 18, 19, 22, 23, 26, 27, 32 |
| | 1 | 500 | 01, 02, 04, 11, 20, 21, 24, 25, 33 |
| | Fig. 17 $C_1$ | | 03, 08, 09, 12, 13, 15, 28, 29, 30, 31 |
| D | Fig. 17 $C_2$ | | 05, 06, 07, 10, 14, 16, 17, 18, 19, 22, 23, 26, 27, 32 |
| | Fig. 17 $C_3$ | | 01, 02, 04, 11, 20, 21, 24, 25, 33 |

## IV. CASE STUDY

Taking an actual WF topology as the example [16], the layout of the WF is depicted in Fig. 4. The WF is composed of thirty-three WTs with the same type (type-4) and capacity. The system parameters are summarized in Table IV, where $P_{m0}$ indicates the steady-state point of each WT. It can be inferred that three factors have impacts on the DVC modes of WT converters, which are the steady-state operating point, the distributed structure of the collector network, and the DVC parameters, $k_{p1}$ and $k_{i1}$. Four cases are set to evaluate the impacts, and further to verify the corresponding DEM.

### *A. Case A: Impact of Steady-State Point and Network*

The steady-state points are set to be different while the DVC parameters are the same to eliminate the effect of control parameter differences in this case. Performing modal calculation based on the constructed linearized WF model, the DVC modal distribution is shown in Fig. 5. It is found that all the modes are located closely. Taking $C$ = 1 and 2 yields a single cluster and $\{\lambda^1, \lambda^2\}$, respectively. The feature vector of each WT is depicted in Fig. 6 (a) and (b), from which it can be seen that the feature vectors of individual WTs are approximately the same. In other words, the WTs even with different operating points and geographical positions share similar DVC dynamic characteristics, as demonstrated in Fig. 7. Therefore, a single-machine representation DEM is suitable in this case. A time-domain simulation is conducted to verify the frequency-domain result, where a sag (5%) of the grid voltage occurs. Fig. 8 illustrates the two coincident response curves. The above results reveal that the impact of steady-state points and networks on the DEM is ignorable.

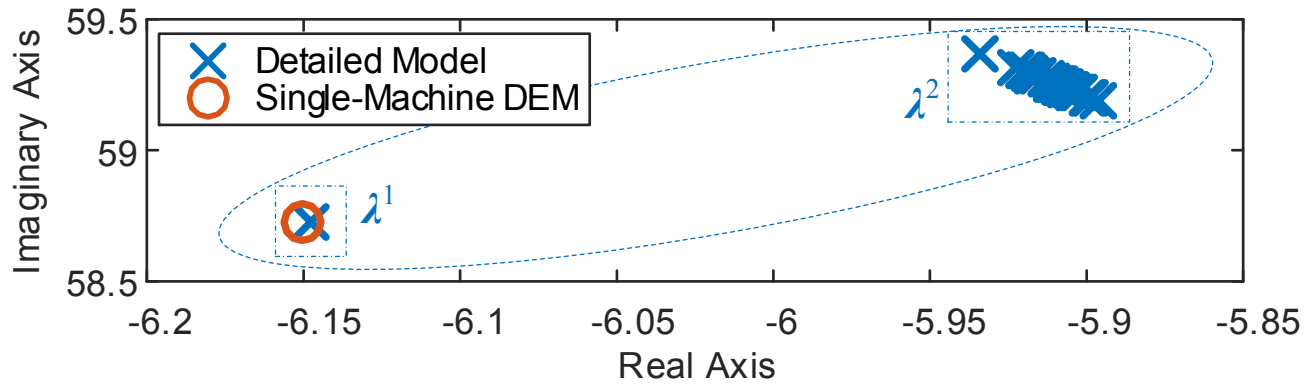

Fig. 5. Modal distributions of the detailed model and DEM in Case A.

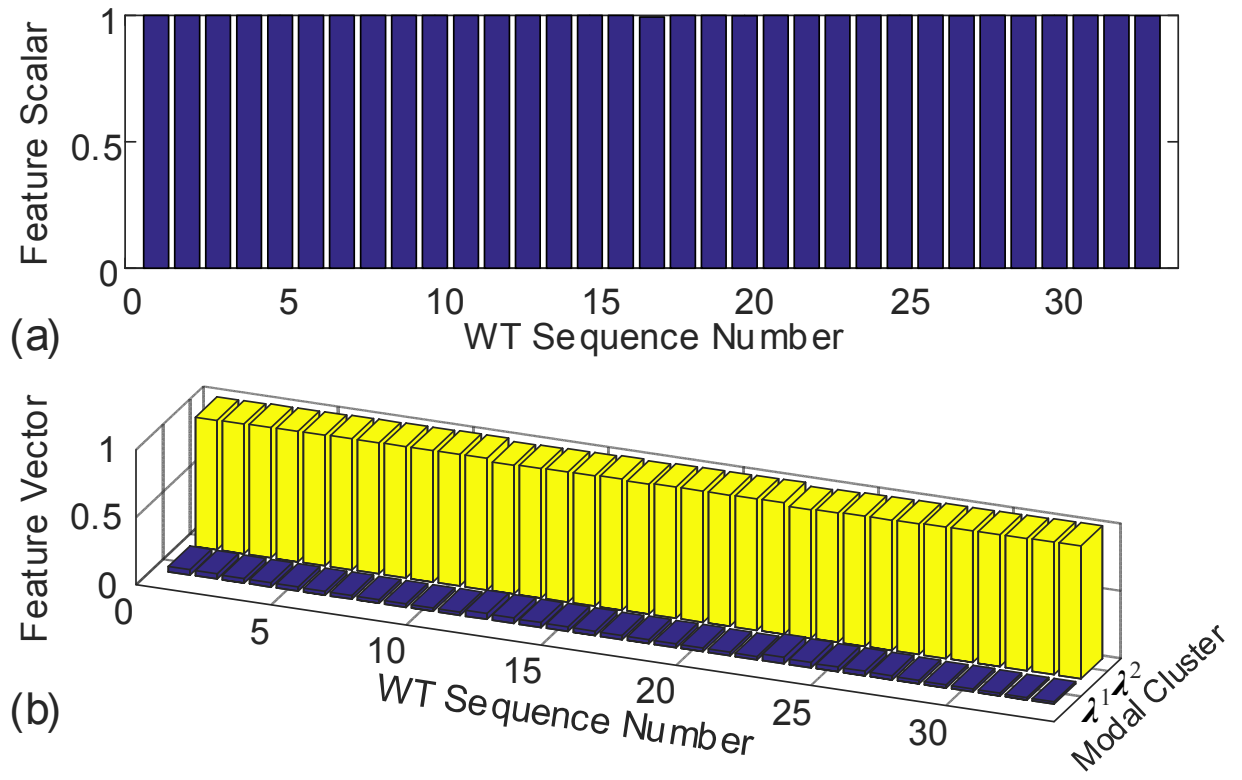

Fig. 6. Feature vector of each WT in Case A. (a) A single modal cluster. (b) Two modal clusters.

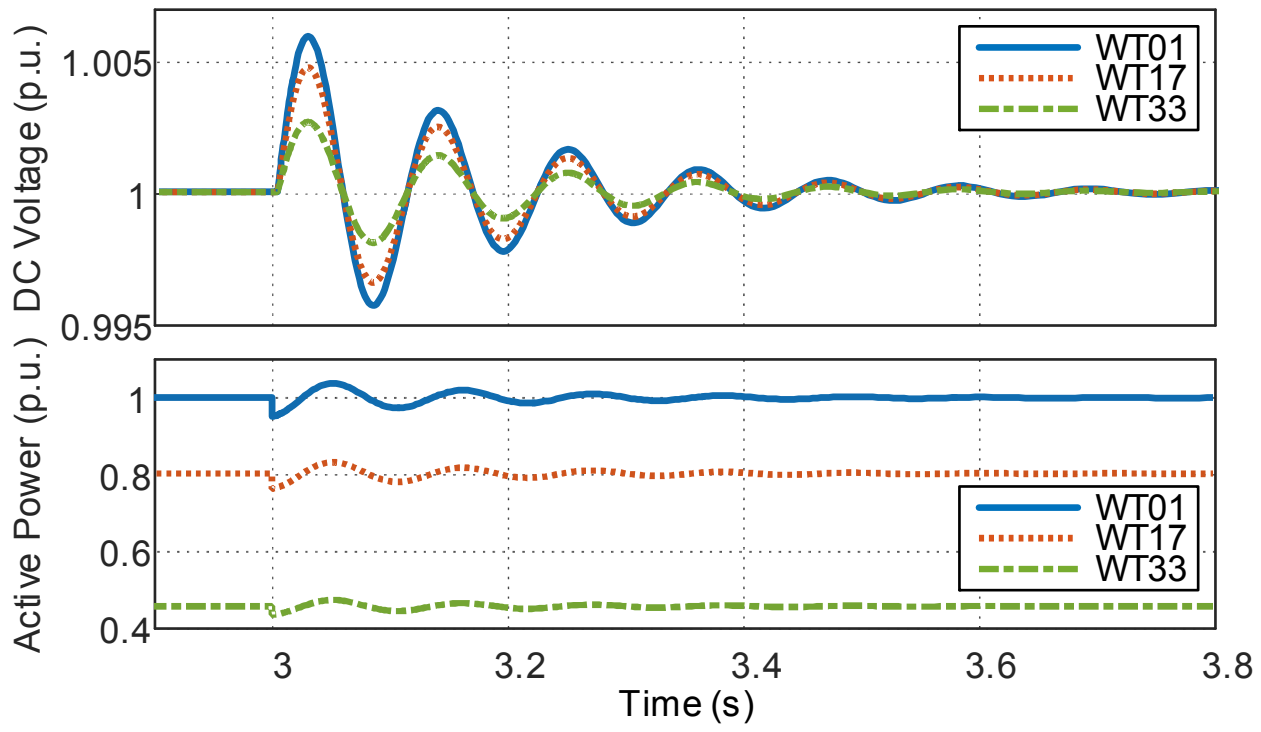

Fig. 7. Responses of several WT examples in Case A.

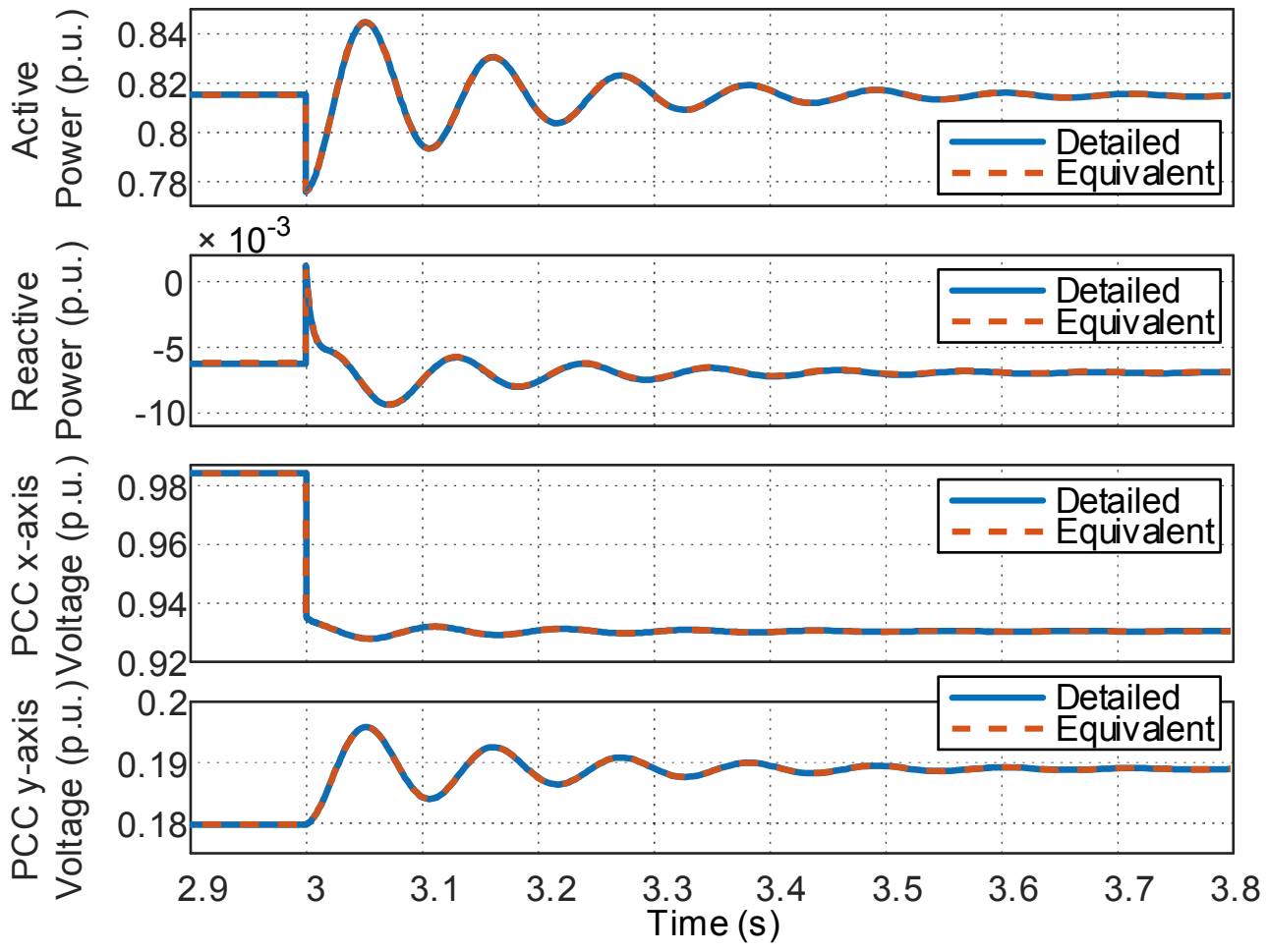

Fig. 8. Response of the detailed model and DEM in Case A.

### B. Case B: Impact of DVC Parameter $k_{p1}$

The DVC proportional parameters of individual WTs are different whereas the integral parameters are the same in this case. From Table V, it can be observed that there are three groups of proportional parameters, which results in three DVC modal clusters $\{\boldsymbol{\lambda}^1, \boldsymbol{\lambda}^2, \boldsymbol{\lambda}^3\}$ (see Fig. 9). Fig. 10 illustrates the feature vectors. The three modal clusters share similar oscillation frequencies but different damping ratios (see Fig. 11). It is indicated in Fig. 10 that WTs with similar motion features should be aggregated into a single machine. Fig. 12 indicates that the presented three-machine DEM provides higher precision than the convention single-machine DEM.

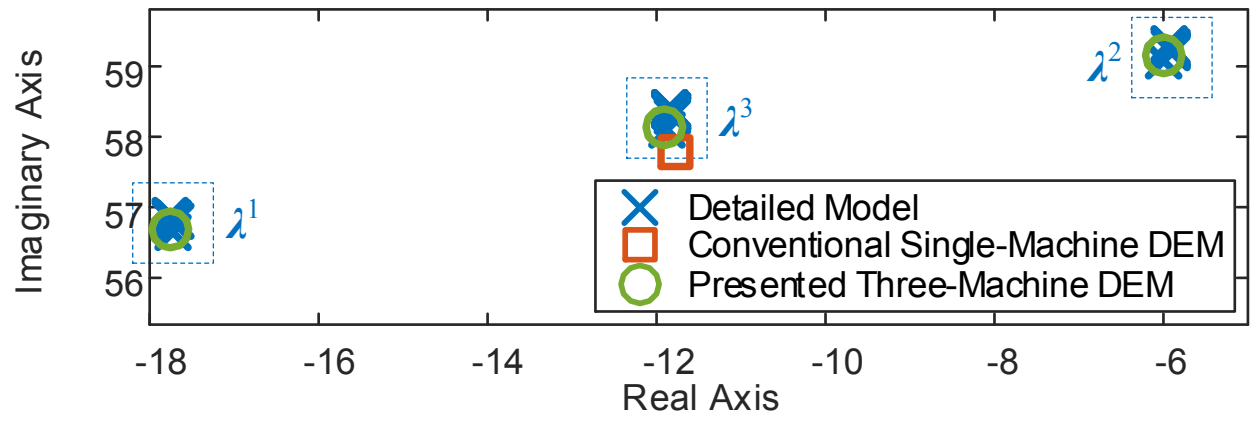

Fig. 9. Modal distributions of the detailed model and DEMs in Case B.

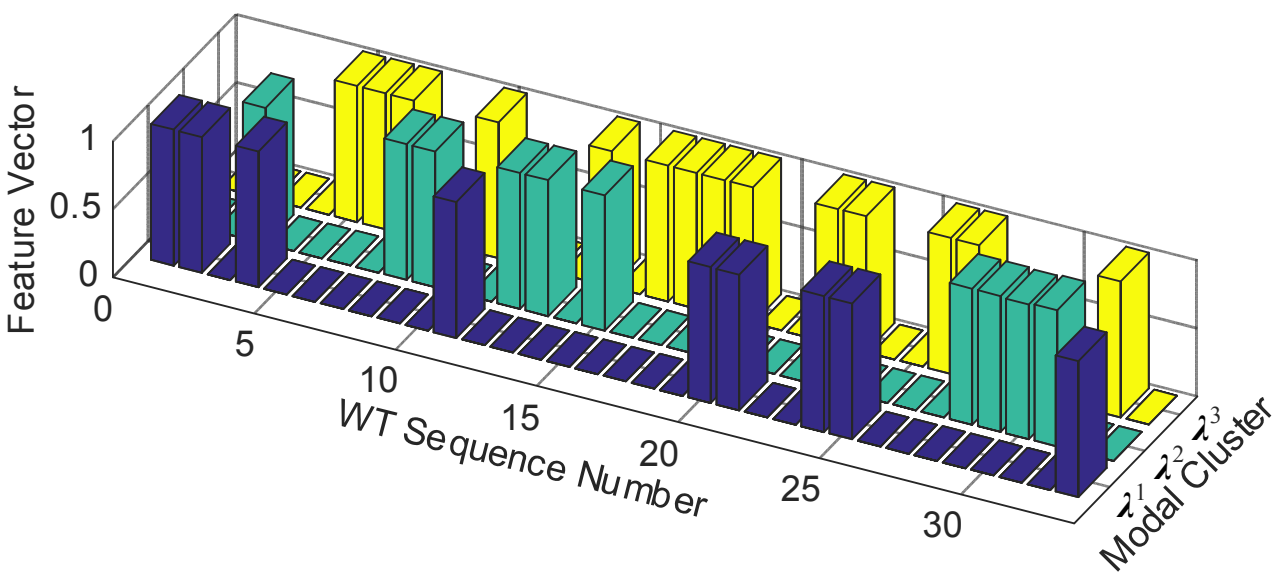

Fig. 10. Feature vector of each WT in Case B.

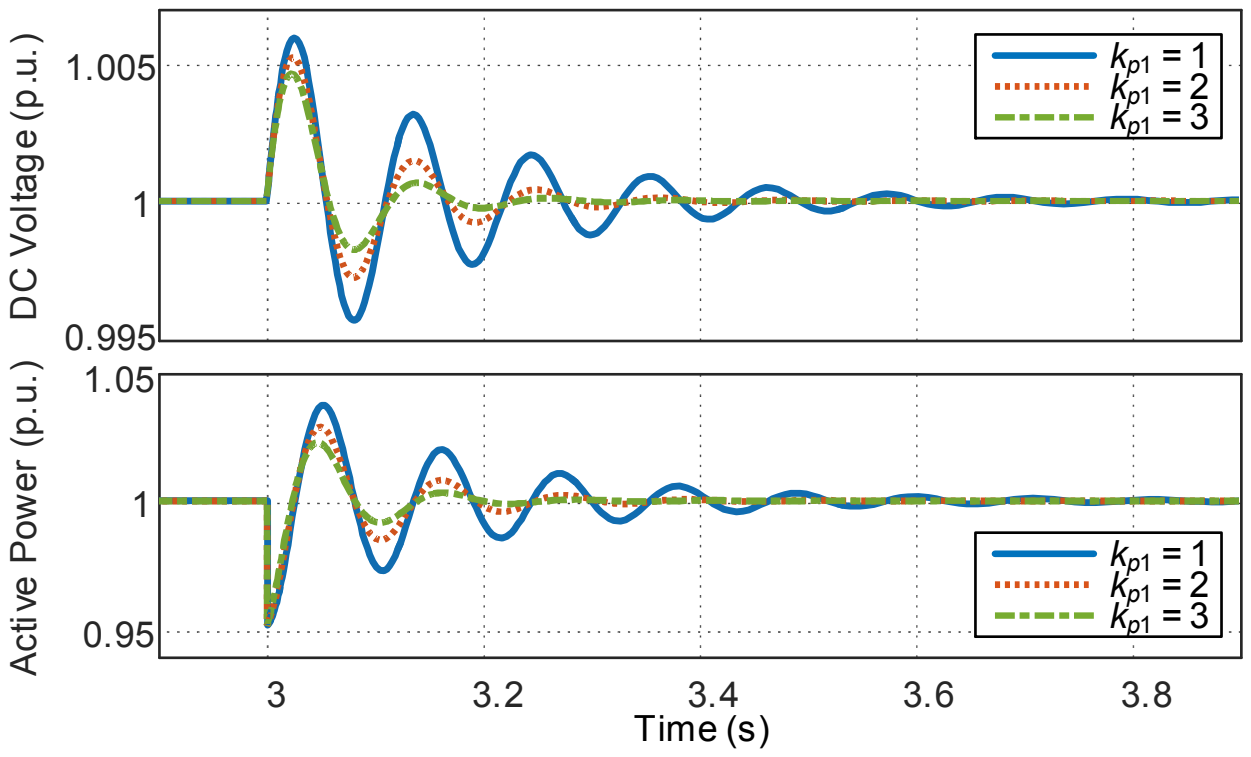

Fig. 11. Responses of several WT examples in Case B.

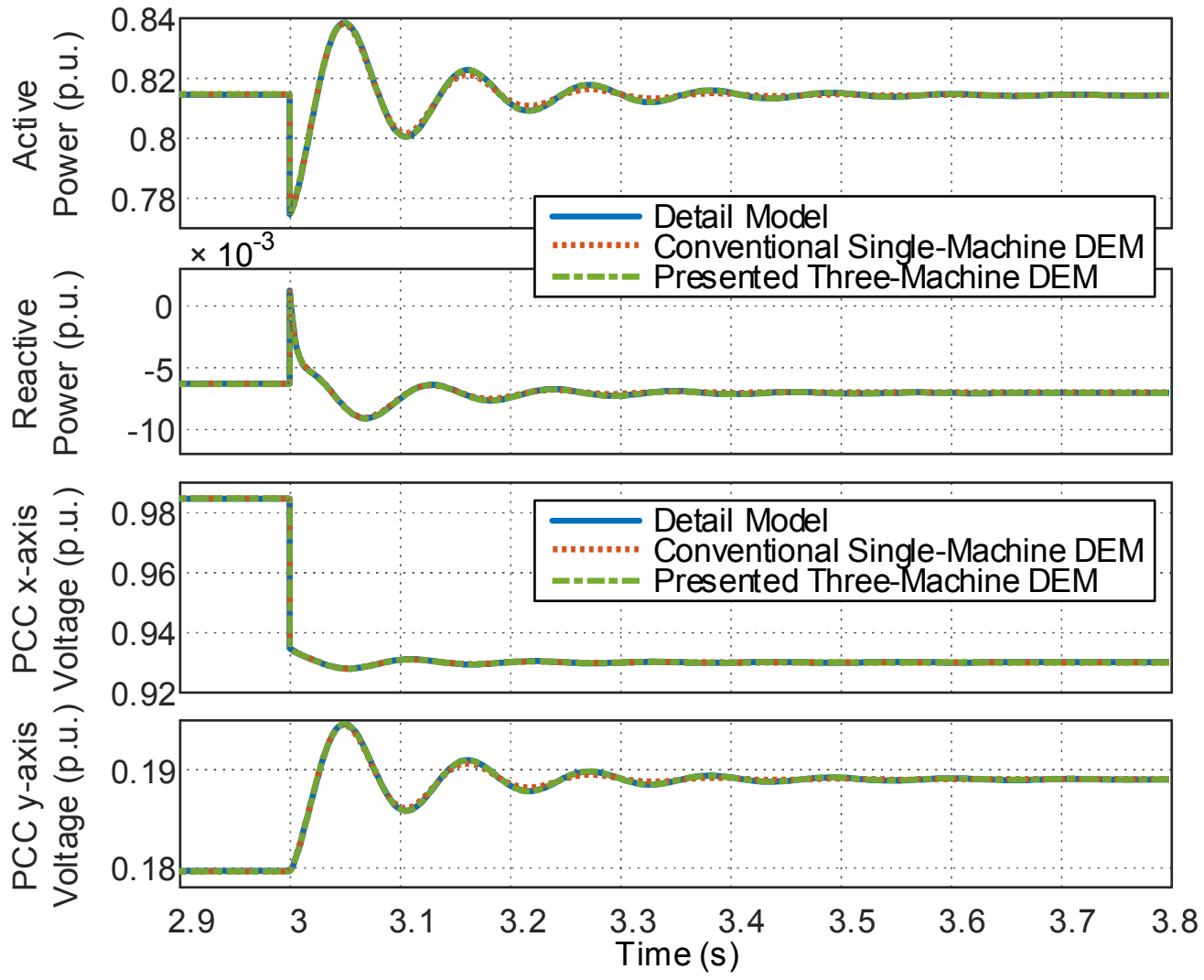

Fig. 12. Responses of the detailed model and DEMs in Case B.

### C. Case C: Impact of DVC Parameter $k_{i1}$

The DVC integral parameters of individual WTs are different whereas the proportional parameters are the same in this case. There are three groups of integral parameters in Table V. The frequency- and time-domain results are illustrated in Figs. 13–16. It can be seen that the WTs with the same DVC parameters exhibit the same DVC dynamic performance, and therefore can be represented by a single WT. All three types of modes are conserved in the three-machine DEM. However, there is only a single DVC mode in the single-machine DEM, therefore leading to quite large modeling errors.

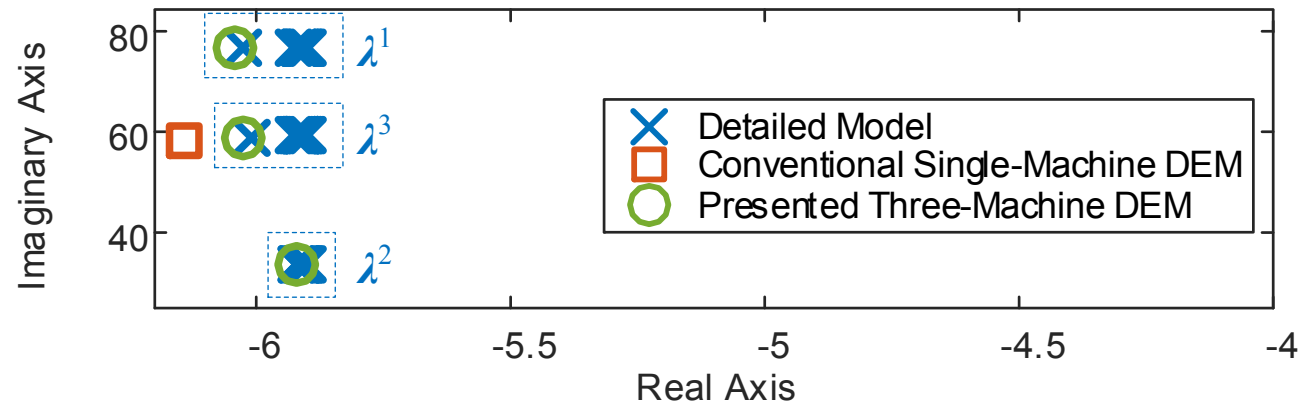

Fig. 13. Modal distributions of the detailed model and DEMs in Case C.

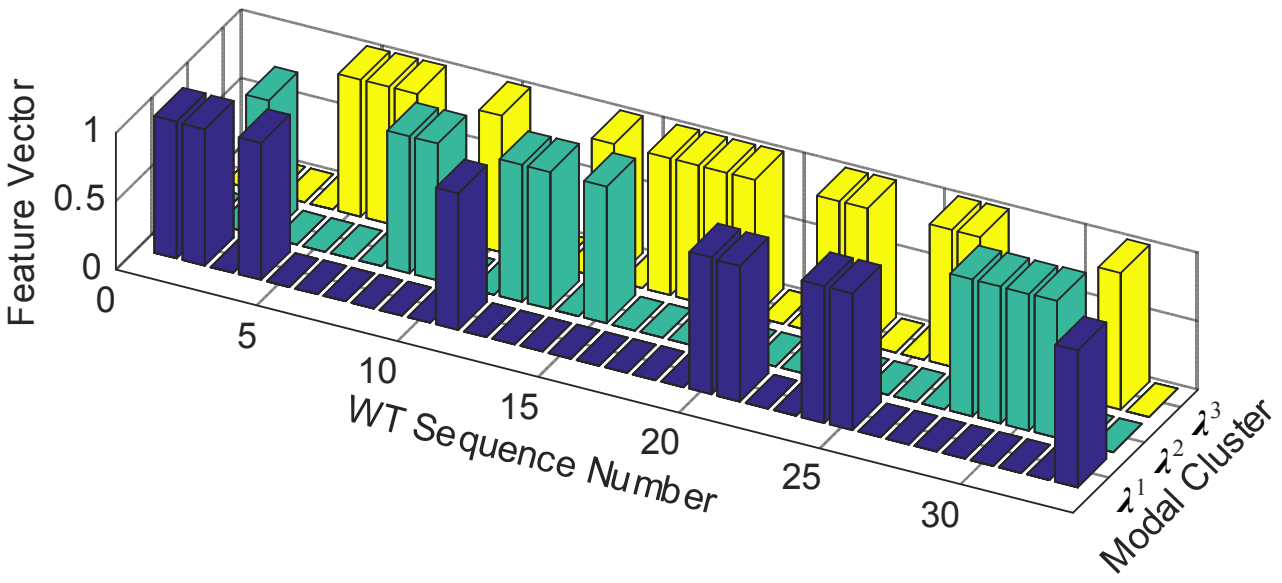

Fig. 14. Feature vector of each WT in Case C.

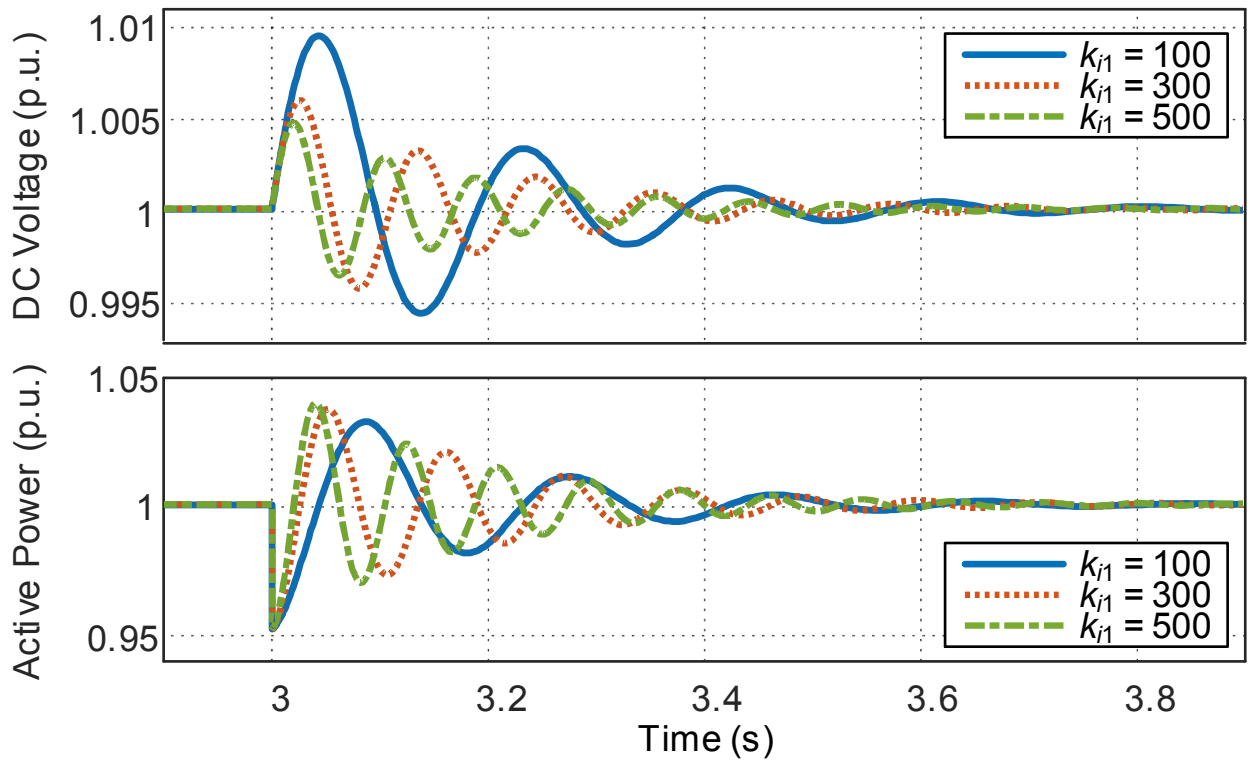

Fig. 15. Responses of several WT examples in Case C.

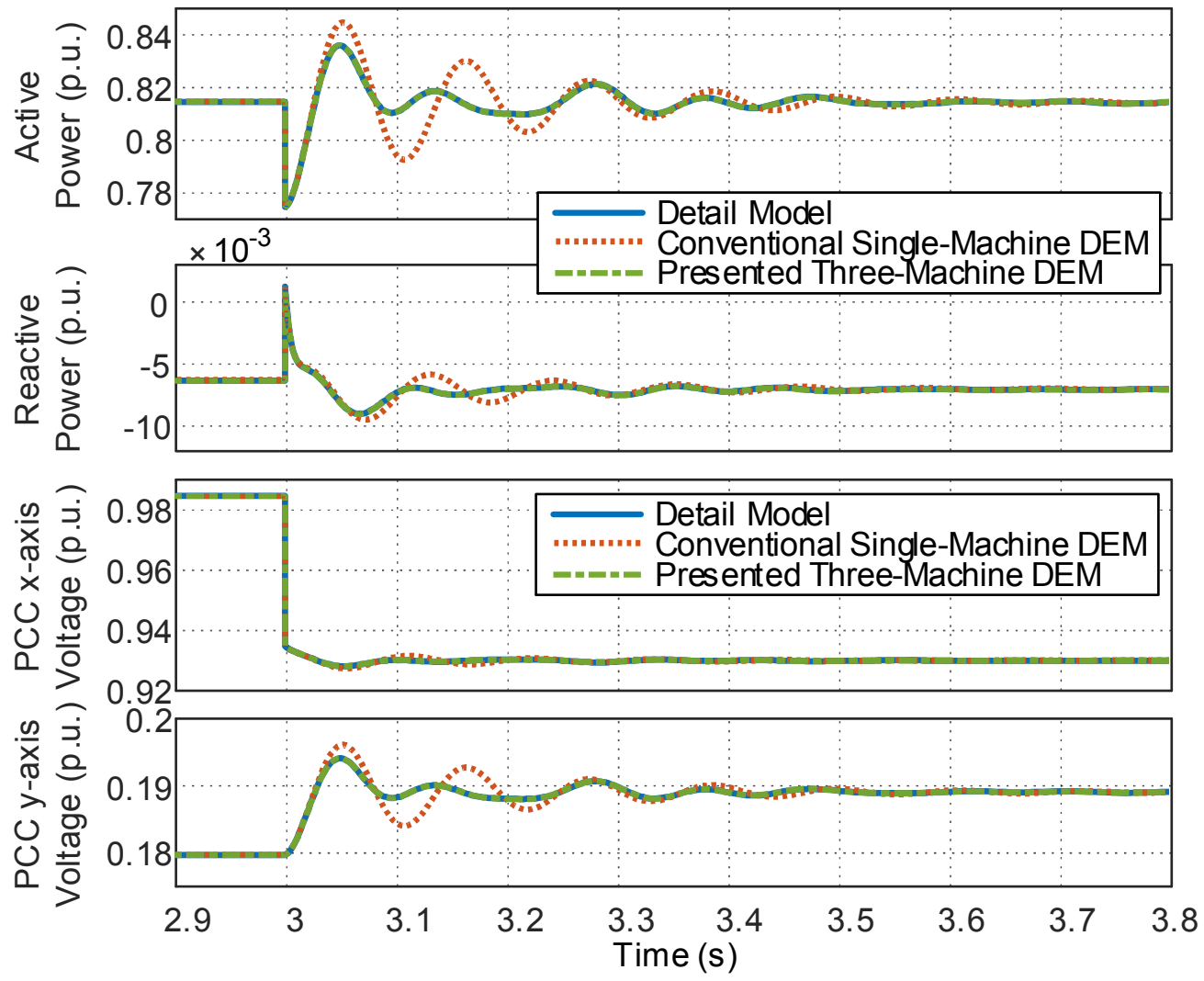

Fig. 16. Responses of the detailed model and DEMs in Case C.

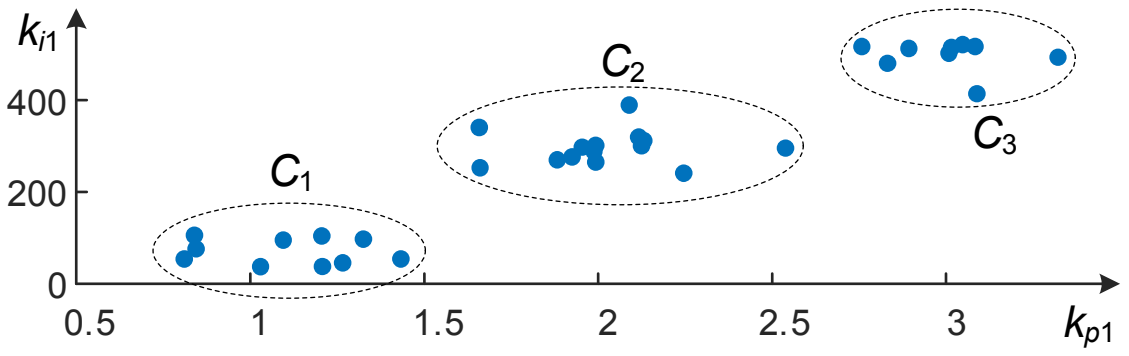

Fig. 17. DVC parameter distribution in Case D.

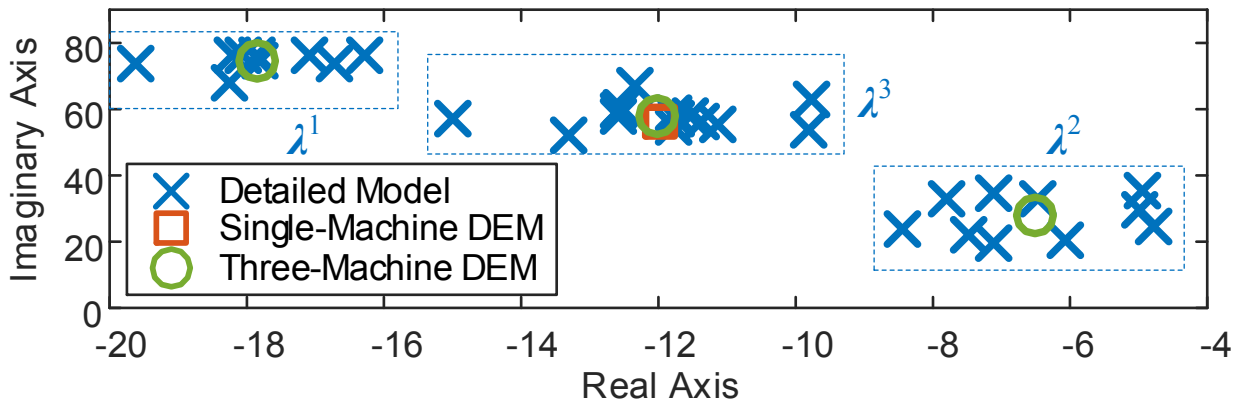

Fig. 18. Modal distributions of the detailed model and DEMs in Case D.

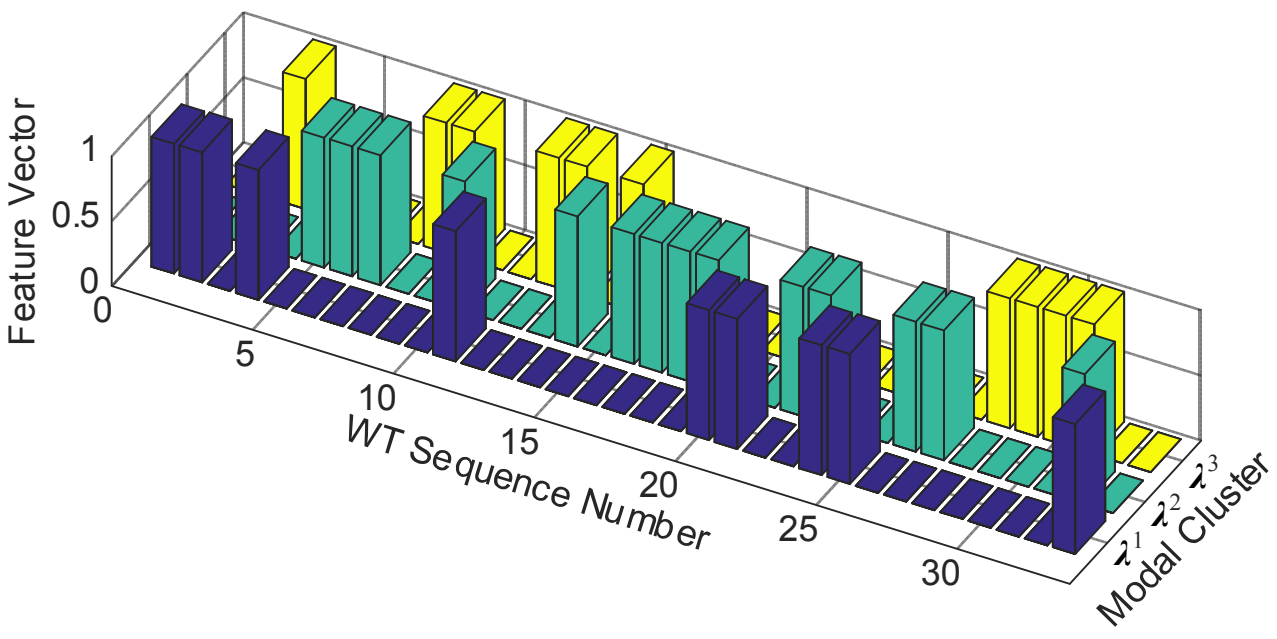

Fig. 19. Feature vector of each WT in Case D.

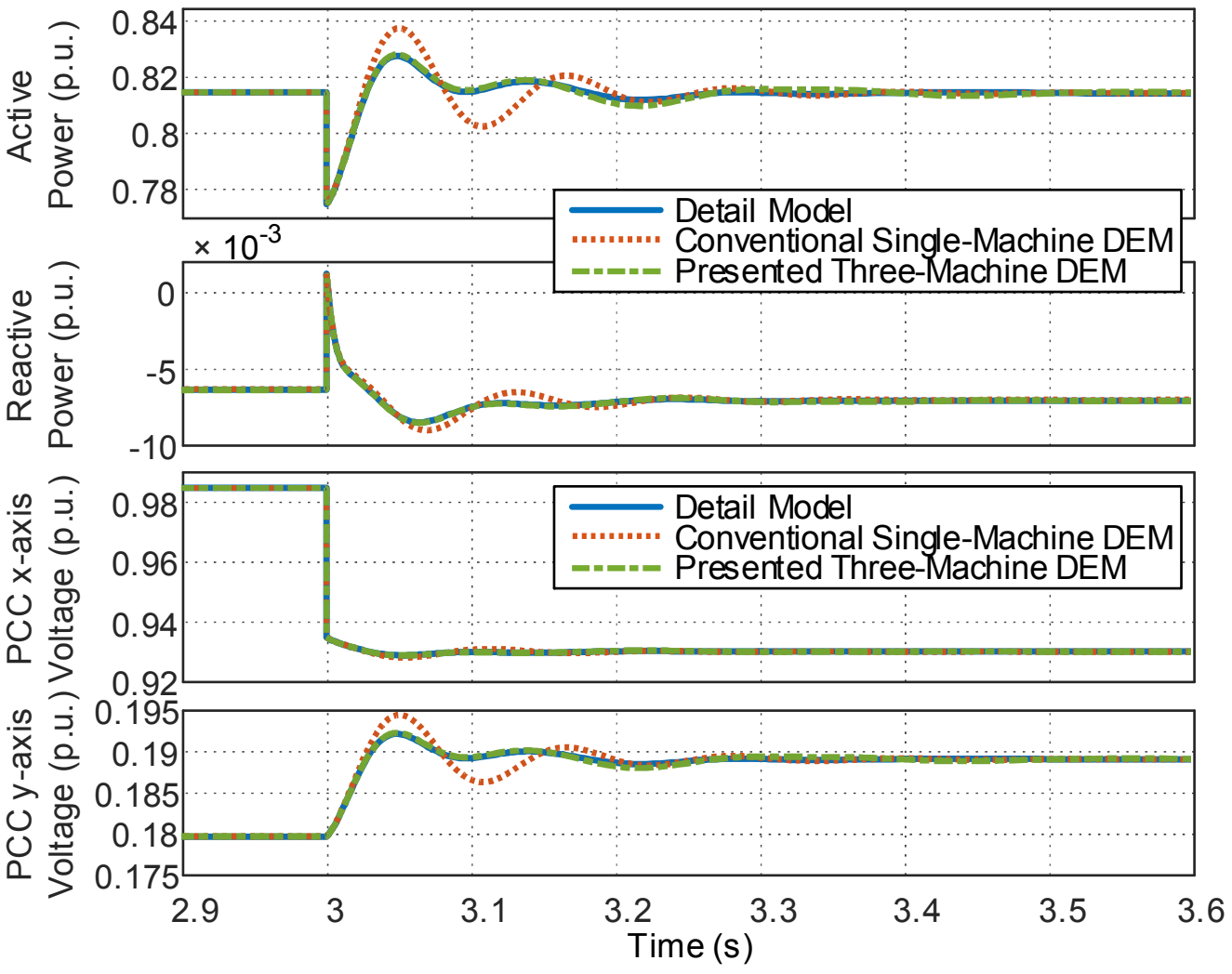

Fig. 20. Responses of the detailed model and DEMs in Case D.

TABLE VI
MODELING ERRORS

| Case | A | | B | | C | | D | |
|---|---|---|---|---|---|---|---|---|
| $C$ | 1 | 2 | 1 | 3 | 1 | 3 | 1 | 3 |
| $E$ (%) | 0.96 | 0.19 | 10.6 | 0.43 | 65.7 | 0.39 | 159.6 | 36.0 |
| $E'$ (%) | 1.14 | 1.14 | 10.2 | 0.49 | 71.1 | 0.45 | 172.7 | 38.6 |

### D. *Case D: Impact of Dispersion of DVC Parameters*

This case is to evaluate the impact of dispersion of DVC parameters, where the parameters of each group are not completely identical (see Fig. 17). The frequency- and time-domain results are shown in Figs. 18–20. Despite the parameter differences, the three-machine DEM is able to display much high accuracy, and it is also more accurate than the single-machine DEM.

The quantitative errors calculated by (14), (15) are summarized in Table VI, where $E$ and $E'$ are approximately identical. Obviously, the errors of the single-machine DEMs ($C = 1$) are generally larger than those of the three-machine DEMs ($C = 3$). Moreover, as revealed by Case D, the dispersion of control parameters certainly increases the error of the DEM to represent the detailed model.

### E. *Discussions*

The foregoing analysis reveals that a single-machine representation is adequate for the DVC dynamic studies, as long as the DVC parameters of individual WTs are identical or similar enough. However, if the parameters are significantly different, multi-machine representations are able to provide much better accuracy than the single-machine representation. The proposed modeling method together with the modeling process provides an effective way to achieve a satisfactory DEM. Actually, the DEM is often a tradeoff between the numbers of representing machines and the accuracy of the DEM.

It is noteworthy that the case study in this section is limited to the DVC mode due to the space limitation. When it comes to other modes of concern, the modeling process should be conducted case by case. Even so, the modeling method is indeed generic.

## V. CONCLUSIONS

High-order and multi-time-scale detailed WF models, composed of all individual WT models and the collector network model, are not applicable for the power system SSS analysis, time-domain simulations, and so forth. A sufficiently simplified yet adequately accurate small-signal DEM is necessary. As for a general form of distributed LTI systems, a generic dynamic equivalent modeling method is firstly proposed. Grounded on the generic method, a concrete dynamic equivalent modeling method of WFs for the small-signal stability analysis is presented. The method has been verified by both frequency- and time-domain example results.

The method can be applied in the following aspects, including but not limited to: 1) derive a DEM to represent the entire WF to analyze the interaction between the WF and power system; 2) represent a large-scale WF with several necessary equivalent machines in real-time simulations to simulate different types, capacities, and parameters of machines; 3) evaluate the dispersion and aggregation of concerned modes within a WF due to the effects of operating points, collector network, and control parameters. Compared with existing methods, which often use time-domain state variables as the clustering index of individual WTs, the proposed method can specifically focus on the concerned modes specified by users. Also, it can provide quantitative errors on the frequency-domain mode clustering, which indicates the accuracy of the DEMs.

Several WFs rather than just a single one are usually integrated to the same grid-connection point, in which different types of WTs, different control strategies, and different control parameters are contained. Future work will address the dynamic equivalent modeling issues on multiple WFs. Furthermore, how to accommodate the uncertainty of parameters and configuration information within WFs using measurements and identification at the POI should draw more attention.